\documentclass[11pt]{article}
\usepackage[cp1251]{inputenc}
\usepackage[T1,T2A]{fontenc}
\usepackage{textcomp}
\usepackage[russian,english]{babel}
\usepackage[centertags]{amsmath}
\usepackage{amsfonts}
\usepackage{amssymb}
\usepackage[pdftex]{hyperref}
\usepackage{graphicx}
\usepackage{graphbox}
\usepackage[numbers,sort&compress]{natbib}

\usepackage{paperinitial}

\paperinitialization{15mm}{15mm}{15mm}{15mm}{2pt}{10pt}

\newcommand{\lan}{\langle}
\newcommand{\ran}{\rangle}

\newcommand{\vk}{\varkappa}
\newcommand{\s}{\sigma}

\newcommand{\be}{\beta}
\newcommand{\ga}{\gamma}

\newcommand{\de}{\delta}

\newcommand{\la}{\lambda}

\newcommand{\spx}{\mathbf{x}}

\newcommand{\spe}{\mathbf{e}}

\begin{document}
\allowdisplaybreaks[4]
\frenchspacing
\setlength{\unitlength}{1pt}
\selectlanguage{english}

\title{{\Large\textbf{Planar wiggler as a tool for generating hard twisted photons}}}

\date{}

\author{O.V. Bogdanov${}^{1),2)}$\thanks{E-mail: \texttt{bov@tpu.ru}},\; P.O. Kazinski${}^{1)}$\thanks{E-mail: \texttt{kpo@phys.tsu.ru}},\; and G.Yu. Lazarenko${}^{1),2)}$\thanks{E-mail: \texttt{laz@phys.tsu.ru}}\\[0.5em]
{\normalsize ${}^{1)}$ Physics Faculty, Tomsk State University, Tomsk 634050, Russia}\\
{\normalsize ${}^{2)}$ Division for Mathematics and Computer Sciences,}\\
{\normalsize Tomsk Polytechnic University, Tomsk 634050, Russia}}

\maketitle

\begin{abstract}

Simple formulas for the probability of radiation of twisted photons by scalar and Dirac particles with quantum recoil taken into account are derived. We show that the quantum recoil does not spoil the selection rule for the forward radiation of twisted photons in the planar undulator: $m+n$ is an even number, where $n$ is the harmonic number and $m$ is the projection of the total angular momentum of the radiated twisted photon. The explicit formulas for the radiation probability of twisted photons produced in the planar wiggler are obtained with account for the quantum recoil. The radiation of twisted photons by GeV electrons in the planar wiggler and in the crystalline undulator is investigated.

\end{abstract}

\section{Introduction}

A rigorous definition of twisted photons in QED is as follows: the twisted photons are the quanta of the electromagnetic field with the definite energy $k_0$, the momentum projection $k_3$, the projection of the total angular momentum $m$ on the same axis, and the helicity $s$. We will call the axis that appears in the definition of twisted photons as the detector axis. The mass-shell condition for such photons reads as $k_{0}=(k_{3}^{2}+k_{\bot}^{2})^{1/2}$. In the paraxial approximation, $n_{\bot}:=k_{\bot}/k_0\ll1$, the projection of the orbital angular momentum $l$ is defined. It is related to the helicity and the total angular momentum as $l=m-s$. By its definition, the projection of angular momentum changes under the shifts of the axis with respect to which it is defined. In performing such a shift, a photon state with the definite projection of the total angular momentum $m$ passes into a superposition of twisted states with all the possible projections $m$ \cite{shift}.

Due to peculiar properties of twisted photons stemming from the fact that they are quanta with the definite projection of angular momentum $m$, the twisted electromagnetic waves provide new instruments for study and solution of fundamental and technical problems. For example, the twisted photons were used in telecommunication to increase the capacity of a channel by employing the projection of angular momentum as an additional quantum number that carries information \cite{2^m}. In microscopy, the use of twisted photons resulted in overcoming the diffraction limit \cite{diflim}. In astrophysics, the twisted photons were used to perform a high-contrast coronagraph \cite{astro}. The optical tweezers based on twisted photons were employed to manipulate nanoparticles \cite{trapp,Greir}. The peculiarities of interaction of twisted photons with atoms were investigated in many papers (see, e.g., \cite{KnyzSerb,MHSSF,SMFSAS14,SGACSSK}). As for interaction of twisted photons with nuclei, we refer to the works \cite{AfSeSol18,TaHaKa}.

The simplest means to create twisted electromagnetic wave is to convert an ordinary plane-wave radiation to a twisted one by employing the holographic or phase plates \cite{KnyzSerb}. However, this approach is unapplicable for production of hard twisted photons in the x-ray and gamma spectral ranges. The pioneering theoretical proposals for generation of hard twisted photons were based on the use of inverse Compton scattering \cite{JenSerprl,JenSerepj,CLHK}, its nonrelativistic limit \cite{TaHaKa,TairKato}, and channeling \cite{ABKT,EpJaZo}. The charges moving along helical trajectories provide a pure source of twisted photons with definite projection of the total angular momentum \cite{SasMcNu,KatohPRL,Hemsing14hel,BKL2,BKL3,BKL4,AfanMikh,BordKN}. This observation was confirmed experimentally in the radiation of helical undulators \cite{BHKMSS,Hemsing14hel,KatohSRexp}. There are other ways to produce hard twisted photons. For example, they can be generated in irradiating a plasma by intense laser beams \cite{ZYCWS18} and in the transition and Vavilov-Cherenkov radiations \cite{BKL5}. Recently, modifying the well-known Baier-Katkov method \cite{BaiKat1,BaKaStrbook}, the general theory of radiation of twisted photons with account for the quantum recoil was developed for both scalar and Dirac particles \cite{BKL4}. In particular, it was shown there that MeV twisted photons can be generated by $180$ GeV electrons in the helical wiggler and by $51.1$ MeV electrons evolving in the laser wave produced by the free-electron laser with photon energy $1$ keV. Sufficiently strong electromagnetic fields of CO$_2$ and Ti:Sa lasers can be employed for production of keV twisted photons by $256$ MeV electrons.

Modern detectors of twisted photons have a rather compact form \cite{RGMMSCFR1, RGMMSCFR2} and are applicable in a quite wide spectral range. The impressive achievements were reached in the methods of sorting electromagnetic radiation by the states with definite projection of the total angular momentum. Nowadays, they allow one to discriminate the projections of angular momentum within the range $|l_{max}|\leqslant30$ \cite{Walsh}. Such a wide resolution of the twisted photon detectors can be used for detailed analysis of the matter structure by twisted photons. It should be noted, however, that the design of twisted photon sorters in the x-ray and gamma ranges is still an open problem \cite{Paroli}. As for the x-ray photons with definite projection of the orbital angular momentum, a triangle aperture can be used as the simplest detector \cite{aperture,taira2019}.

\section{Constant energy approximation}

We use in this paper the system of units such that $\hbar=1$, $c=1$, and $e^{2} = 4 \pi \alpha $, where $\alpha\approx 1/137$ is the fine structure constant. To describe the radiation of twisted photons, it is convenient to introduce the basis
\begin{equation}\label{spin_eigv}
    \mathbf{e}_\pm:=\mathbf{e}_1\pm i\mathbf{e}_2,\quad \mathbf{e}_3,
\end{equation}
where $\{\spe_1,\spe_2,\spe_3\}$ is a right-handed orthonormal triple and $\spe_3$ is directed along the detector axis. Any vector can be written as
\begin{equation}
    \mathbf{r}= r_{3} \mathbf{e}_{3} +\frac{1}{2}\big( r_{+} \mathbf{e}_{-} + r_{-} \mathbf{e}_{+} \big), \quad r_{\pm}=(\mathbf{r},\mathbf{e}_\pm),\quad r_{3}=(\mathbf{x},\mathbf{e}_3).
\end{equation}
In the paper \cite{BKL4}, the formulas were obtained for the probability of radiation of twisted photons by relativistic charged particles with the quantum recoil taken into account. These formulas are the analog of semiclassical formulas for the probability of radiation of plane-wave photons with the quantum recoil \cite{BaiKat1,BaKaStrbook}. For the charged scalar particle, we have
\begin{equation}\label{prob_by_scal}
    dP(s,m,k_\perp,k_3)=e^2 \Big|\int_{-\infty}^\infty dt e^{-i[k_0-k_\perp^2/(2P_{0 i})] q_{i} t+i q_{i} k_3x_{3}}
    q^{1/2} (t) \big(\tfrac12\big[\dot{x}_{+} a_- +\dot{x}_{-} a_+ \big] +\dot{x}_{3} a_3 \big)\Big|^2 \Big(\frac{k_\perp}{2k_{0}}\Big)^{3}\frac{dk_3dk_\perp}{2\pi^2}.
\end{equation}
where $\spx(t)$ is found by solving the Lorentz equations for a charged particle in the given electromagnetic field and $t$ is the laboratory time. The trajectory $\spx(t)$ describes approximately the motion of the center of particle's wave-packet. Also
\begin{equation}
    P_0(t):=m_e\ga(t),\qquad P'_{0}(t):=P_{0}(t)-k_0,\qquad q(t):=P_0(t)/P'_{0}(t),
\end{equation}
where $m_e$ is the electron mass and $\ga(t)$ is the Lorentz factor. Besides, $q_{i}=P_{0 i}/P'_{0 i}=const $, where $P_{0 i}$ is the initial energy of the particle. The notation has been introduced:
\begin{equation}\label{mode_func_an}
\begin{split}
   a_3&\equiv a_3(m,k_3,k_\perp;\spx)=J_{m} (k_{\bot} |x_{+}|) e^{i m \arg( x_{+})}=:j_m(k_\perp x_+,k_\perp x_-),\\
    a_{\pm}&\equiv a_{\pm} (s,m,k_\perp; \spx)=\frac{i k_{\bot}}{s k_{0} \pm k_{3}} j_{m\pm1}(k_\perp x_+,k_\perp x_-).
\end{split}
\end{equation}
As far as the Dirac fermions are concerned, the formula is written as
\begin{equation}\label{prob_by_Dir}
\begin{split}
    dP(s,m,k_\perp,k_3) & =e^2\int_{-\infty}^\infty \frac{dt_1dt_2}{4P'_{01} P'_{02}} e^{-i(k_0-k_\perp^2/(2P_{0 i}))q_{i}(t_2-t_1)+ik_3q_{i}(x_{23}-x_{13})}\times\\
    &\times\Big\{(P_{01}+P'_{01})(P_{02}+P'_{02})(\tfrac12[\dot{x}_{1-} a^*_- +\dot{x}_{1+} a^*_+]+\dot{x}_{13}a^*_3) (\tfrac12[\dot{x}_{2+} a_- +\dot{x}_{2-} a_+]+\dot{x}_{23}a_3)+\\
    &+\frac{k_0^2}{4}\big[(\dot{x}_{1+}a^*_+-in_\perp a^*_+(m-1)) (\dot{x}_{2-}a_+ +in_\perp a_+(m-1))+ \\
    &+(\dot{x}_{1-} a^*_-+in_\perp a^*_-(m+1)) (\dot{x}_{2+}a_--in_\perp a_-(m+1)) \big]\Big\}\Big(\frac{k_\perp}{2k_{0}}\Big)^{3}\frac{dk_3dk_\perp}{2\pi^2},
\end{split}
\end{equation}
where $n_\perp:=k_\perp/k_0$, $n_3:=(1-n_\perp^2)^{1/2}$, and, for brevity, we denote
\begin{equation}\label{a_pm3}
\begin{gathered}
    a_\pm\equiv a_\pm(s,m,k_3,k_\perp; \spx_2 ),\qquad a_3\equiv a_3 (s,m,k_\perp; \spx_2),\\
    a^*_\pm\equiv a^*_\pm(s,m,k_3,k_\perp; \spx_1),\qquad a^*_3\equiv a^*_3(s,m,k_\perp; \spx_1).
\end{gathered}
\end{equation}
In formula \eqref{prob_by_Dir}, only those arguments of the mode functions are indicated that differ from \eqref{a_pm3} and $\spx_{1,2}:=\spx(t_{1,2})$. These formulas were derived in \cite{BKL4} under the assumptions
\begin{equation}\label{estimates}
    k_0/P_0\lesssim1,\qquad |\dot{x}_\pm|\sim \vk/\gamma,\qquad|\dot{x}_3|\approx1,\qquad|\dot{x}_3-n_3|\lesssim\vk^2/\gamma^2,\qquad |n_\perp|\lesssim \vk/\gamma,\qquad n_3\approx1,
\end{equation}
where $\vk=\max(1,K)$, $K=\lan\be_\perp\ran\ga$ is the undulator strength parameter \cite{Bord.1, BKL2}. It is that region of parameters where the main part of radiation of a relativistic particle is concentrated.

Formulas \eqref{prob_by_scal}, \eqref{prob_by_Dir} simplify in the case when the energy of a particle, $P_{0}(t)$, is constant or its change is negligible on the radiation formation scale. In the leading order in $\vk/\ga$, the expression entering into the radiation amplitude for a scalar particle \eqref{prob_by_scal} becomes
\begin{equation}
    \tfrac12\big[\dot{x}_{+} a_- +\dot{x}_{-} a_+ \big] +\dot{x}_{3} a_3 \approx j_{m}+ \frac{s i \dot{x}_{s}}  {n_{\bot}} j_{m-s}.
\end{equation}
Then formula \eqref{prob_by_scal} for a scalar particle is reduced to
\begin{equation}\label{prob_by_scal2}
    dP(s,m,k_\perp,k_3)= q_{i} e^{2}  \Big|\int dt e^{-i (k_{0}-k_{\bot}^{2}/(2P_{0 i}))q_{i} t+i k_{3} q_{i} x_{3}} \big( a_{3} + \tfrac12 \dot{x}_{s} a_{-s}\big)  \Big|^{2} \Big(\frac{k_{\bot}}{2 k_{0}}\Big)^{3} \frac{dk_{3} dk_{\bot}}{2 \pi^2}.
\end{equation}
In the case of a Dirac particle, the spin contributions have the form
\begin{equation}
    \dot{x}_{\mp}a_{\pm} \pm in_\perp a_{\pm}(m\mp 1) \approx  (s\mp 1)\Big( \frac{i \dot{x}_{\mp} }{n_{\bot}}  j_{m\pm1} \mp  j_{m} \Big),
\end{equation}
in the leading order in $\vk/\ga$. Only one of the two terms survives for any given value of the helicity $s$. As a result, the sum of these contributions can be written as
\begin{equation}\label{dPa}
    dP_{a}(s,m,k_\perp,k_3)=e^2 \frac{(1-q_{i})^{2}}{4 }\Big|\int{dt e^{-i(k_0-k_\perp^2/(2P_{0 i}))q_{i} t+ik_3 q_i x_{3}} \big( a_{3} + \tfrac12 \dot{x}_{s} a_{-s} \big)} \Big|^{2} \Big(\frac{k_\perp}{2k_{0}}\Big)^{3}\frac{dk_3dk_\perp}{2\pi^2}.
\end{equation}
The main contribution to \eqref{prob_by_Dir} is obtained from \eqref{prob_by_scal2} by replacing the common factor $q_i$ by $(1+q_{i})^{2}/4$. Adding the spin contributions \eqref{dPa} to the main one, we arrive at
\begin{equation}\label{prob_by_Dir2}
    dP(s,m,k_\perp,k_3)= \frac{1+q_{i}^{2}}{2} e^{2} \Big|\int{dt e^{-i (k_{0}-k_{\bot}^{2}/(2P_{0 i}))q_{i} t+i k_{3} q_{i} x_{3}} \big( a_{3} + \tfrac12 \dot{x}_{s} a_{-s}\big) } \Big|^{2} \Big(\frac{k_{\bot}}{2 k_{0}}\Big)^{3} \frac{dk_{3} dk_{\bot}}{2 \pi^2}.
\end{equation}
The validity of the derived formulas is confirmed by the particular examples considered in \cite{BKL4,BKL2,Matveev}. We see that, in this approximation, the radiation probability is a modulus squared of the radiation amplitude taking into account the quantum recoil. This amplitude can be used describe the coherent effects in the radiation of beams of charged particles \cite{BKb,BKLb,BKL5} caused by their transverse structure.

\section{Planar Wiggler}

As an application of the above formulas, let us consider the radiation of a charged particle in a planar undulator or in a laser wave with linear polarization. In those cases, the charged particle moves along the $z$ axis uniformly and rectilinearly with the velocity $\be_\parallel\approx1$ for $t<-TN/2$ and $t>TN/2$. As for $t\in[-TN/2,TN/2]$, the particle evolves along the trajectory (see, e.g, \cite{Bord.1})
\begin{equation}\label{planar_wiggl}
    x_{3} = \beta_{\parallel} t -\frac{\beta_{\parallel} K^{2}}{4 \omega \ga^{2}}\sin(2\omega t) , \quad x_{1}=0 , \quad x_{2}= \frac{\sqrt{2}\be_\parallel K}{\ga\omega}\sin(\omega t),
\end{equation}
where $t$ is the laboratory time, $T=2\pi/\omega$ is the oscillation period, and $N\gg1$ is the number of periods (the number of sections in the undulator). In fact, the second term in $x_3(t)$ can be neglected under the assumptions \eqref{estimates}. The trajectory is joined continuously at the end-points of the interval $[-TN/2,TN/2]$. The undulators in the wiggler regime correspond to $K\gtrsim 1$ and $K/\ga\ll1$ (see for details, e.g., \cite{Bord.1}).

Performing the calculations along the lines of \cite{BKL2}, where the radiation of twisted photons in planar undulators without the quantum recoil was considered, we deduce the radiation amplitude
\begin{equation}\label{A}
    A=\sum_{n=1}^{+\infty}{\delta_{N}\big[k_{0} q_{i}(1- k_0 n_{\bot}^{2}/(2 P_{0 i}))-q_{i} k_{3}  \beta_{\parallel}-\omega n\big]  \Big(f_{n,m}-\frac{K}{\sqrt{2}\ga n_{\bot} }\Big(f_{n+1,m-s}+f_{n-1,m-s}\Big)\Big)},
\end{equation}
where
\begin{equation}
    \delta_{N}[x_{n}] := \int_{-TN/2}^{+TN/2}{\frac{dt}{2\pi}e^{-i tx_{n}}}=\frac{\sin(TN x_{n}/2)}{\pi x_{n}},
\end{equation}
and
\begin{equation}\label{f}
    f_{n,m}=\pi (1+ (-1)^{n+m})\sum_{k=-\infty}^{+\infty}{ (-1)^{k} J_{k}\Big( q_i \frac{\beta_{\parallel} K^{2} k_{0}}{4 \omega \ga^{2}}\Big)J_{(n-m)/2-k}\Big(\frac{\beta_{\parallel} K k_{\bot}}{\sqrt{2} \omega \ga}\Big)J_{(n+m)/2-k}\Big(\frac{\beta_{\parallel} K k_{\bot}}{\sqrt{2} \omega \ga}\Big)}.
\end{equation}
As in the case of radiation without the quantum recoil \cite{BKL2}, it follows from formulas \eqref{f} and \eqref{A} that the selection rule $m+n$ is an even number is obeyed.

In the general case, the energy spectrum of radiated photons following from \eqref{A} has a rather awkward form \cite{BKL4}. Nevertheless, neglecting the term $k_{0} n_{\bot}^{2}/(2 P_{0})$ in the argument of the delta function, we come to
\begin{equation}\label{k_0}
    k_0=\frac{n\omega}{1-n_3 \beta_\parallel +n\omega/P_{0 i}}\;\Leftrightarrow\;\frac{1}{k_0}=\frac{1}{\bar{k}_0}+\frac{1}{P_{0 i}},
\end{equation}
where $\bar{k}_{0}$ is the energy of radiated photons without the quantum recoil. It is seen from this expression that the energy of radiated photons cannot exceed the energy of a radiating particle.

\begin{figure}[!t]
\centering
\includegraphics*[width=0.6\linewidth]{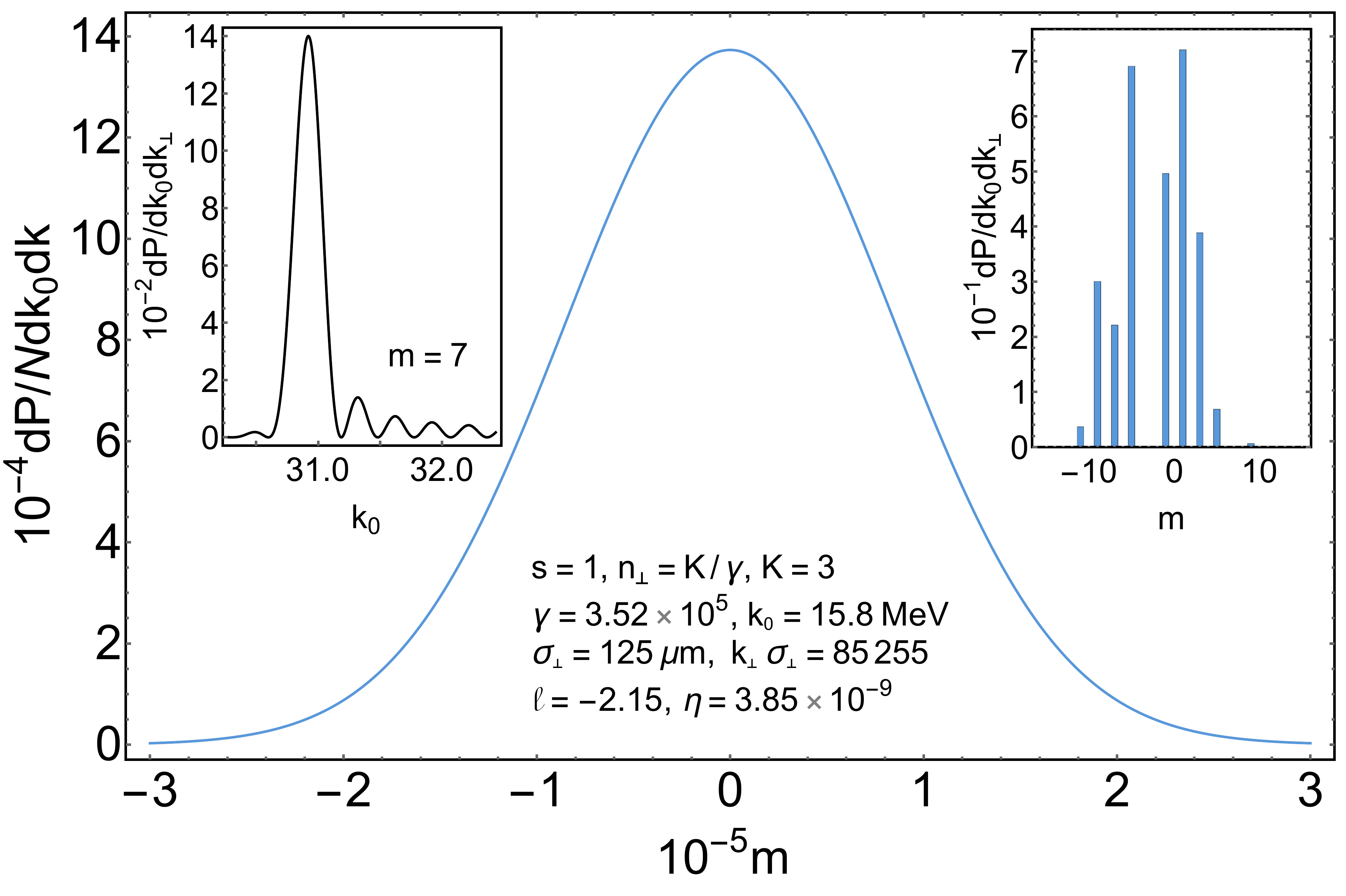}
\caption{{\footnotesize The forward radiation of twisted photons by $180$ GeV unpolarized electrons in the planar wiggler. The seventh harmonic is presented. The wiggler period is $0.72$ cm, the number of periods $N=15$, and the magnetic field strength in the wiggler is $63$ kG. The quantum recoil reduces the energy of radiated photons by $1.38$ keV in comparison with the classical formula. The ratio of probabilities \eqref{prob_by_Dir2} and \eqref{prob_by_scal2} calculated by the use of formulas \eqref{prob_by_scal}, \eqref{prob_by_Dir} is constant over the whole domain of the variables $k_0$ and $m$. It is equal to $(1+q_i^2)/(2q_i)=:1+\eta$. The main plot: The distribution over $m$ of incoherently radiated twisted photons by the axially symmetric Gaussian bunch of particles with the transverse size $\s_\perp=125$ $\mu$m (see the details in \cite{BKb,BKLb}). The parameter $\ell$ is the projection of the total angular momentum of radiated photons per photon. It coincides with the same quantity for the radiation by one particle (right inset) in accordance with the theorem proved in \cite{BKb}. The left inset: The probability to record a twisted photon produced by electrons in the wiggler against the photon energy. The photon energy is measured in the electron rest energies. The right inset: The probability distribution over $m$ of radiated twisted photons by the electron moving along the center of the bunch. The selection rule $m+n$ is an even number is fulfilled for $n=7$.}}
\label{planar_ond_plots}
\end{figure}

As a result, the probability of radiation of a twisted photon by a charged scalar particle is
\begin{equation}\label{PS}
\begin{split}
    dP(s,m,k_\perp,k_3)&= q_{i} e^{2} \delta_{N}^2\big[k_{0} q_{i}(1- k_0 n_{\bot}^{2}/(2 P_{0 i}))-q_{i} k_{3}  \beta_{\parallel}-\omega n\big]\times\\
    &\times\Big|f_{n,m}-\frac{K}{\sqrt{2}\ga n_{\bot} }\Big(f_{n+1,m-s}+f_{n-1,m-s}\Big) \Big|^{2} \Big(\frac{k_{\bot}}{2 k_{0}}\Big)^{3} \frac{dk_{3} dk_{\bot}}{2 \pi^2}.
\end{split}
\end{equation}
As for the Dirac particle, we find
\begin{equation}\label{PD}
\begin{split}
    dP(s,m,k_\perp,k_3)&= \frac{1+q_{i}^{2}}{2} e^{2} \delta_{N}^2\big[k_{0} q_{i}(1- k_0 n_{\bot}^{2}/(2 P_{0 i}))-q_{i} k_{3}  \beta_{\parallel}-\omega n\big]\times\\
    &\times\Big|f_{n,m}-\frac{K}{\sqrt{2}\ga n_{\bot} }\Big(f_{n+1,m-s}+f_{n-1,m-s}\Big) \Big|^{2} \Big(\frac{k_{\bot}}{2 k_{0}}\Big)^{3} \frac{dk_{3} dk_{\bot}}{2 \pi^2}.
\end{split}
\end{equation}
The probability of radiation of twisted photons by a beam of relativistic particles can easily be obtained from \eqref{PS}, \eqref{PD} by employing the formulas presented in \cite{BKb,BKLb,BKL5}.

The coherent radiation of helically microbunched beams of charged particles is concentrated at the harmonics \cite{BKLb}
\begin{equation}\label{harmonics_bunch}
    k_0=2\pi\chi n_c \be_3/\de,\qquad \chi n_c>0,\;n_c\in \mathbb{Z},
\end{equation}
stemming from the periodic structure of the beam. Here the undulator radiation of the beam is considered, $n_c$ is the number of the coherent harmonic, $\chi$ specifies the handedness of the bunch, and $\de$ is the helix pitch in the laboratory reference frame. For the coherent radiation of such a beam, the strong addition rule is fulfilled \cite{BKLb}. Namely, the spectrum of twisted photons over $m$ produced at the $n_c$-th coherent harmonic is shifted by $n_c$ with respect to the radiation spectrum of twisted photons over $m$ produced by one particle moving along the beam center. As a result, the projection of the total angular momentum per photon, $\ell_\rho$, can be increased as $\ell_\rho = \ell_1 + \chi n_c$, where $\ell_1$ is the projection of the total angular momentum per photon for the radiation generated by one particle moving along the beam center. The helically microbunched beams of charged particles were already used in experiments for generation of twisted photons with orbital angular momentum \cite{HKDXMHR}. The details of experimental techniques used for imprinting the helical structure on the beam can be found in \cite{HemStuXiZh14}.

\begin{figure}[!t]
\centering
\includegraphics*[width=0.6\linewidth]{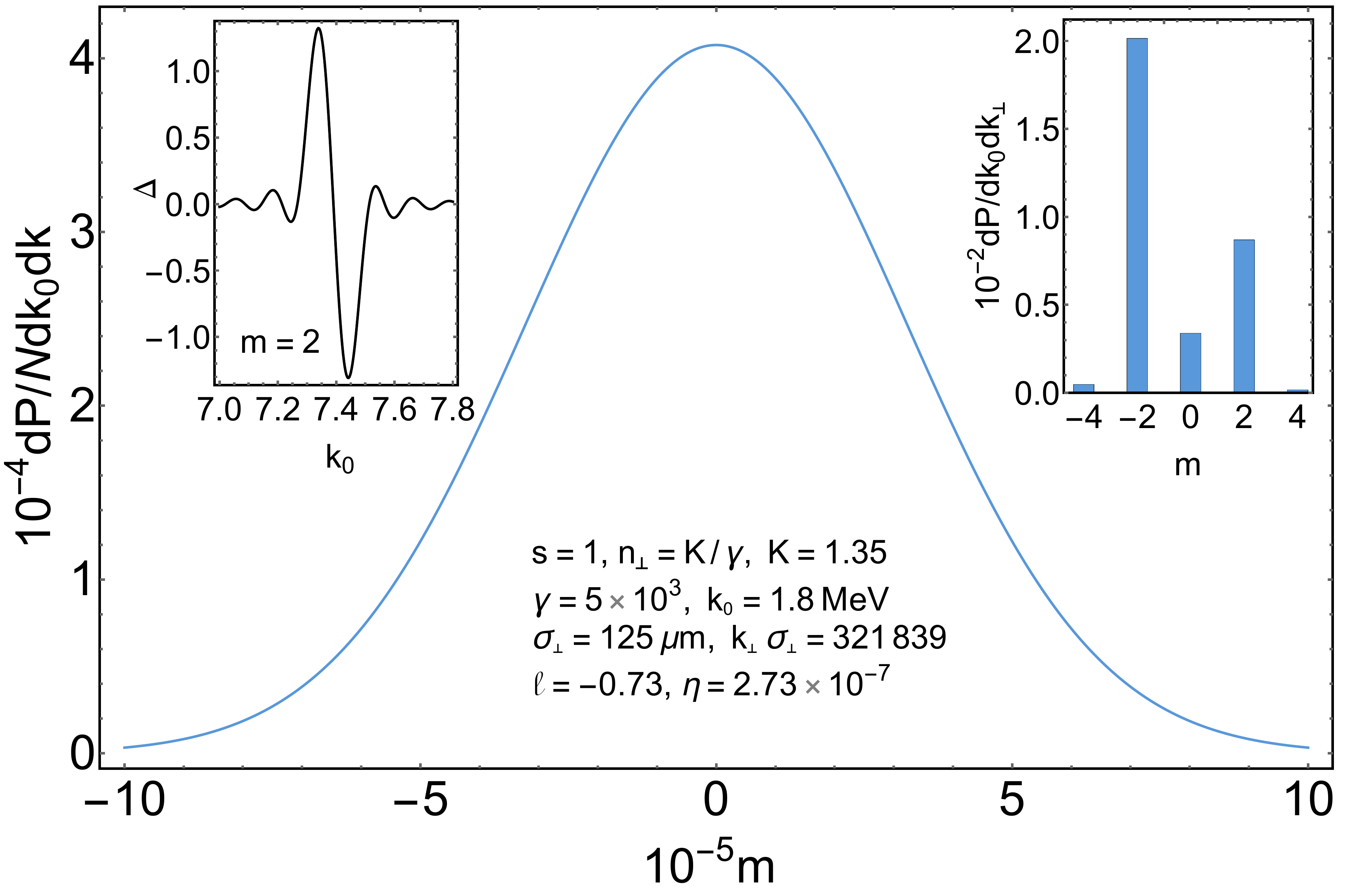}
\caption{{\footnotesize The forward radiation of twisted photons by $2.56$ GeV unpolarized electrons in the crystalline undulator. The second harmonic is presented. The trajectory has the form \eqref{planar_wiggl} with $\omega=2\pi\be_\parallel/\la_0$, the period of crystalline undulator is $\la_0=23$ $\mu$m, the amplitude of oscillations of electron's trajectory is $a=1$ nm. The number of sections is $N=15$. These parameters are taken from \cite{KKSG}. The quantum recoil reduces the energy of radiated photons by $1.39$ keV in comparison with the classical formula. The ratio of probabilities \eqref{prob_by_Dir2} and \eqref{prob_by_scal2} calculated by the use of formulas \eqref{prob_by_scal}, \eqref{prob_by_Dir} is constant over the whole domain of the variables $k_0$ and $m$. It is equal to $(1+q_i^2)/(2q_i)=1+\eta$. The main plot: The distribution over $m$ of incoherently radiated twisted photons by the axially symmetric Gaussian bunch of particles with the transverse size $\s_\perp=125$ $\mu$m. The projection of the total angular momentum of radiated photons per photon $\ell$ coincides with the same quantity for the radiation by one particle. The left inset: The difference between the probability \eqref{prob_by_Dir2} and the same quantity given by the classical formula \cite{BKL2} against the photon energy. The photon energy is measured in the electron rest energies. The right inset: The probability distribution over $m$ of radiated twisted photons by the electron moving along the center of the bunch. The selection rule $m+n$ is an even number is fulfilled for $n=2$.}}
\label{planar_ond_cr_plots}
\end{figure}

As the particular examples we consider the radiation of twisted photons by the planar wiggler Fig. \ref{planar_ond_plots} and crystalline undulator Fig. \ref{planar_ond_cr_plots}. The ratio of the radiation probability for the Dirac particle and the same quantity for the scalar one coincides perfectly with $(1+q_i^2)/(2q_i)$. The selection rule $m+n$ is an even number is satisfied.

\section{Conclusion}

Let us briefly recapitulate the results. Starting from the general formulas for the radiation probability of twisted photons with quantum recoil taken into account that were derived in \cite{BKL4}, we obtained simpler formulas under the assumptions that the radiated photons are paraxial and the energy of a charged particle as it is given by the Lorentz equations can be regarded as a constant. The radiation of both scalar and Dirac charged particles was investigated. As an application of these general formulas, the explicit formulas for the radiation probability of twisted photons by charged particles in planar undulators were deduced. It was proved that the selection rule established in \cite{BKL2} for the radiation of twisted photons by charged particles in planar undulators when the quantum recoil is discarded holds also in the case when this recoil is included into the theory. The numerical simulations for the radiation of twisted photons in the planar wiggler and the crystalline undulator showed a good agreement with the general formulas derived.

\paragraph{Acknowledgments.}
The work is supported by the RFBR grant 20-32-70023.

\end{document}